\documentclass[aps,prb,twocolumn,showpacs,preprintnumbers,amsmath,amssymb,superscriptaddress,notitlepage]{revtex4-2}

\usepackage[colorlinks=true,allcolors=blue]{hyperref}
\usepackage{xcolor}
\usepackage{txfonts}
\usepackage{graphicx}
\usepackage{color}
\usepackage{dcolumn}
\usepackage{bm}
\usepackage{amsmath}
\usepackage{feynmp}
\usepackage{float}
\usepackage{braket}
\usepackage{cleveref}

\newcommand{\nc}{\newcommand}
\nc{\bea}{\begin{eqnarray}}
\nc{\eea}{\end{eqnarray}}
\nc{\vx}{{\mathbf{r}}}
\nc{\vm}{\mathbf{m}}

\begin{document}

\title{Quantum encoding of structured light into in-plane topological spin textures}

\author{Pavel~A.~Vorobyev}
\email{p.vorobyev@unsw.edu.au}
\affiliation{
	School of Physics, The University of New South Wales, Sydney 2052, Australia
}

\author{Daichi~Kurebayashi}
\affiliation{
	School of Physics, The University of New South Wales, Sydney 2052, Australia
}

\author{Oleg~A.~Tretiakov}
\email{o.tretiakov@unsw.edu.au}
\affiliation{
School of Physics, The University of New South Wales, Sydney 2052, Australia
}
\date{\today}

\begin{abstract}

Structured light offers a powerful means of controlling light-matter interactions through multiple tunable optical degrees of freedom. Using micromagnetic simulations, we investigate the nucleation of asymmetric bimerons and antibimerons by pulsed Laguerre-Gaussian optical vortices in chiral ferromagnetic thin films with $C_{nv}$ and $D_{2d}$ symmetries, respectively. For optical vortices with orbital angular momentum (OAM) $|m|=1$, circularly polarized beams deterministically nucleate a single bimeron or antibimeron via the interplay of spin angular momentum, OAM, and magnetic chirality, whereas linearly polarized beams produce textures whose topological charge directly follows the OAM ($Q=m$). Optical vortices with OAM $|m|>1$ nucleate clusters and other configurations composed of multiple spin textures, whose morphology and topological charge depend sensitively on the optical quantum numbers and pulse parameters. These findings reveal a route to topology-selective writing through the encoding of optical quantum numbers into distinct in-plane topological magnetic states.
\end{abstract}

\maketitle

\section{Introduction} 

Topological spin textures~\cite{Nagaosa2013,GobelPhysRep2021}, such as skyrmions~\cite{Skyrme1962, Belavin1975, Bogdanov2002, Muhlbauer2009} and bimerons~\cite{Kharkov2017, Gobel2019, Gao2019, Chen2025}, have emerged as promising candidates for encoding and processing information in nonvolatile memory and logic devices~\cite{Fert2017, Guoqiang2017, Rajib2024}. Their topological robustness, particle-like nature, nanoscale size, and current-driven manipulation make them appealing information carriers~\cite{Fert2013, Sampaio2013} in the pursuit of higher storage density, faster operation, and device miniaturization. Furthermore, the ability to stabilize some of these spin textures in multiple distinct topological states~\cite{Li20, Vorobyev2024} makes them compelling building blocks for multi-valued logic~\cite{Hurst1984, Andreev2022}, enabling enhanced data capacity and energy-efficient operation~\cite{Jo2021}. Realizing the potential of these topological spin textures in practical spintronic devices requires the ability to reliably write, manipulate, and annihilate them on demand.

Writing magnetic skyrmions and other topological textures locally and with minimal energy consumption remains one of the central challenges for spintronic device engineering~\cite{Zhou2025}. Conventional approaches rely on injecting local currents~\cite{Hrabec2017, Buttner2017}, applying magnetic fields to shrink labyrinth domains into skyrmions~\cite{Soumyanarayanan2017, Fallon2020}, exploiting transient Joule heating induced by nanosecond current pulses~\cite{Legrand2017, Lemesh2018}, and utilizing strain-mediated control~\cite{Nii2015, Feng2021}. Despite significant advances, these methods remain subject to limitations that include high current densities~\cite{Zhang_2020}, excessive Joule heating losses~\cite{Litzius2020}, stochastic nucleation~\cite{Kurebayashi2022}, and the need for intricate device architectures~\cite{Jiang2015, Juge2022}. These limitations have motivated growing interest in noncontact approaches for the deterministic and local creation of target spin textures.

\begin{figure}[h!]
	\centering
	\includegraphics[width=1\linewidth]{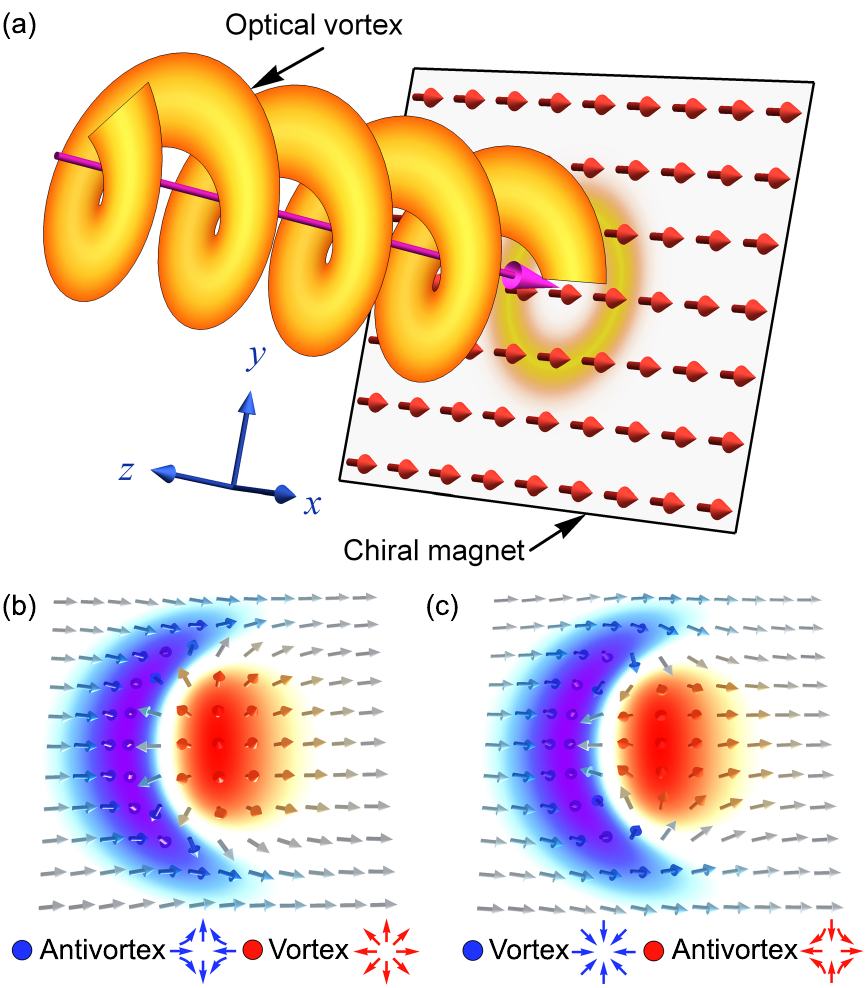}
	\caption{(a) Schematic illustration of the model. An OV beam is incident on a chiral magnetic thin film. The spiral represents the helical phase front of the OV, the magenta arrow indicates its propagation direction, and the red arrows denote the initial uniform in-plane magnetization of the chiral magnet. (b) Magnetic configuration of an asymmetric bimeron. (c) Magnetic configuration of an asymmetric antibimeron. The sketches below panels (b) and (c) show the vortex-antivortex pair forming each spin texture.}
	\label{fig1}
\end{figure}

All-optical control of magnetization~\cite{Lambert2014} provides a noncontact route to the ultrafast generation of topological spin textures. While spatially uniform laser pulses have been shown to nucleate skyrmions and bimerons by transiently modifying magnetic interactions~\cite{Finazzi2013, Koshibae2014, SoongGeun2018, Zhu2026}, structured light offers a far more selective paradigm of optical control because its polarization, azimuthal phase winding, and radial structure are explicitly governed by optical quantum numbers. In particular, optical vortices (OVs)~\cite{Allen1992, Arnold2008, Shen2019} have been theoretically proposed as a powerful means of deterministically generating topological spin textures in magnetic media~\cite{Fujita2017}. To date, however, theoretical studies have focused primarily on out-of-plane magnetized systems, predicting the creation of skyrmions and related out-of-plane textures~\cite{Hiroyuki2017, Guan2023, Zhang2026}. By contrast, the use of structured light to nucleate topological spin textures within in-plane magnetized systems remains largely unexplored.

Here, we demonstrate the encoding of the optical quantum numbers of structured light into in-plane topological magnetic states. Using micromagnetic simulations, we model pulsed Laguerre-Gaussian (LG) OVs as spatially structured magnetic fields coupled to chiral ferromagnetic (FM) thin films through the Zeeman interaction [see Fig.~\ref{fig1}(a)]. For modes with orbital angular momentum (OAM) $|m|=1$, circularly polarized OVs enable the deterministic writing of single asymmetric bimeron [Fig.~\ref{fig1}(b)] and antibimeron [Fig.~\ref{fig1}(c)] textures through the interplay of spin angular momentum (SAM), OAM, and magnetic chirality, whereas linearly polarized OVs directly encode the OAM into the topological charge $Q$ of the nucleated texture, yielding $Q=m$. For modes with $|m|>1$, we demonstrate the nucleation of clusters and other configurations composed of multiple spin textures, with the final state depending sensitively on the optical quantum numbers and pulse parameters. Our results point to structured light as a powerful tool for the topology-selective writing of in-plane spin textures.

\section{Optical vortex micromagnetic model} 

We consider an ultrathin FM film lying in the $xy$-plane. The film is positioned at the focal plane ($z=0$) of a pulsed OV beam propagating along the $z$-direction. The magnetic free energy of the system is given by
\begin{equation}
F=\!\int\! d^3r \left\{A(\mathbf{\nabla} \mathbf{m})^2 + \epsilon_{a} +	\epsilon_{\text{DM}} - M_{s}\mathbf{m}\cdot\left[\mathbf{B}_{\text{st}} +\mathbf{B}(\mathbf{r},t)\right] \right\},
\label{eq1_fe}
\end{equation}
where $A > 0$ is the exchange stiffness constant, $\mathbf{m} = \mathbf{M}/M_{s}$ is the normalized magnetization, $M_{s}$ is the saturation magnetization, and $\epsilon_{a}=K_{x}[1-(\mathbf{m}\cdot  \mathbf{e}_{x})^2]$ is the in-plane easy-axis anisotropy energy density with the effect of magnetostatic interactions being accounted for via a local shape anisotropy contribution to the effective constant $K_x > 0$. We consider two distinct forms of the Dzyaloshinskii-Moriya interaction (DMI) energy density $\epsilon_{\text{DM}}$. For a $C_{n\text{v}}$-symmetric system, which stabilizes asymmetric bimerons~\cite{Li20}, it is expressed as
\begin{equation}
\epsilon_{DM} = D \left[ m_z (\nabla \cdot \mathbf{m}) - (\mathbf{m} \cdot \nabla) m_z \right],
\label{eq:dmi_cnv}
\end{equation}
whereas for a $D_{2d}$-symmetric system, which hosts asymmetric antibimerons~\cite{Vorobyev2024}, it takes the form
\begin{equation}
\epsilon_{DM}=D \left(\mathbf{e}_x \cdot \mathbf{m} \times \partial_y\mathbf{m} + \mathbf{e}_y \cdot \mathbf{m} \times \partial_x\mathbf{m}\right),
\label{eq:dmi_d2d}
\end{equation}
where $D$ is the DMI constant, whose sign determines the preferred magnetic chirality. The static external magnetic field $\mathbf{B}_{\text{st}} = B_{x} \mathbf{e}_{x}$ is applied along the easy axis, and $\mathbf{B}(\mathbf{r},t)$ denotes the time-dependent magnetic field of the OV pulse. In the initial state, the film is uniformly magnetized along the $+x$-direction. In our micromagnetic simulations, we adopt the following material parameters: $M_{s} = 111\text{ kA m}^{-1}$, $A = 0.6\text{ pJ m}^{-1}$, $D = 0.13\text{ mJ m}^{-2}$, and $K_x = 8.5\text{ kJ m}^{-3}$, together with a static magnetic field $B_x = 40\text{ mT}$.

To model the structured optical excitation, we represent the OV magnetic field using the focal-plane profiles of LG modes within the paraxial approximation~\cite{Allen1992}. Within our model, the magnetic component of the OV couples to the magnetization through the Zeeman term in Eq.~\eqref{eq1_fe} \footnote{Potentially, the electric-field component of the OV may also couple to the magnetization through material-dependent magnetoelectric or spin-orbit mediated mechanisms. In systems with sufficiently strong magnetoelectric response, this coupling could provide an excitation channel comparable to, or stronger than, the direct Zeeman coupling considered here.}. At the focal plane ($z=0$), its spatial profile is written as \begin{equation}
\label{eq:lg_mode}
\mathbf{B}_{m,p}(\rho, \phi) =  B_0 \left( \frac{\rho}{W} \right)^{|m|}  e^{-\frac{\rho^2}{W^2} + i m \phi} L_p^{|m|} \left( \frac{2 \rho^2}{W^2} \right) \mathbf{e}_s,
\end{equation}
where $(\rho,\phi)$ are polar coordinates in the film plane, 
$B_0$ is the amplitude coefficient of the OV magnetic field, $W$ is the effective size (waist) of the beam, and $L_p^{|m|}$ are the generalized Laguerre polynomials. The vector $\mathbf{e}_s$ represents the polarization state of the OV, with the index $s$ labeling the polarization-dependent SAM of the optical field. Here, $s=0$ denotes linear polarization, which carries no net SAM, whereas $s=\pm1$ denote circular polarizations carrying opposite SAM projections along the propagation direction. In this convention, $s=+1$ and $s=-1$ correspond to left- and right-handed circular polarizations, respectively. The integer $m$ is the azimuthal index that determines the OAM of the mode, with the phase factor $\exp(im\phi)$ describing the azimuthal phase winding, whereas the integer $p$ is the radial index governing the radial mode structure.

\begin{figure}[b]
	\centering
	\includegraphics[width=1\linewidth]{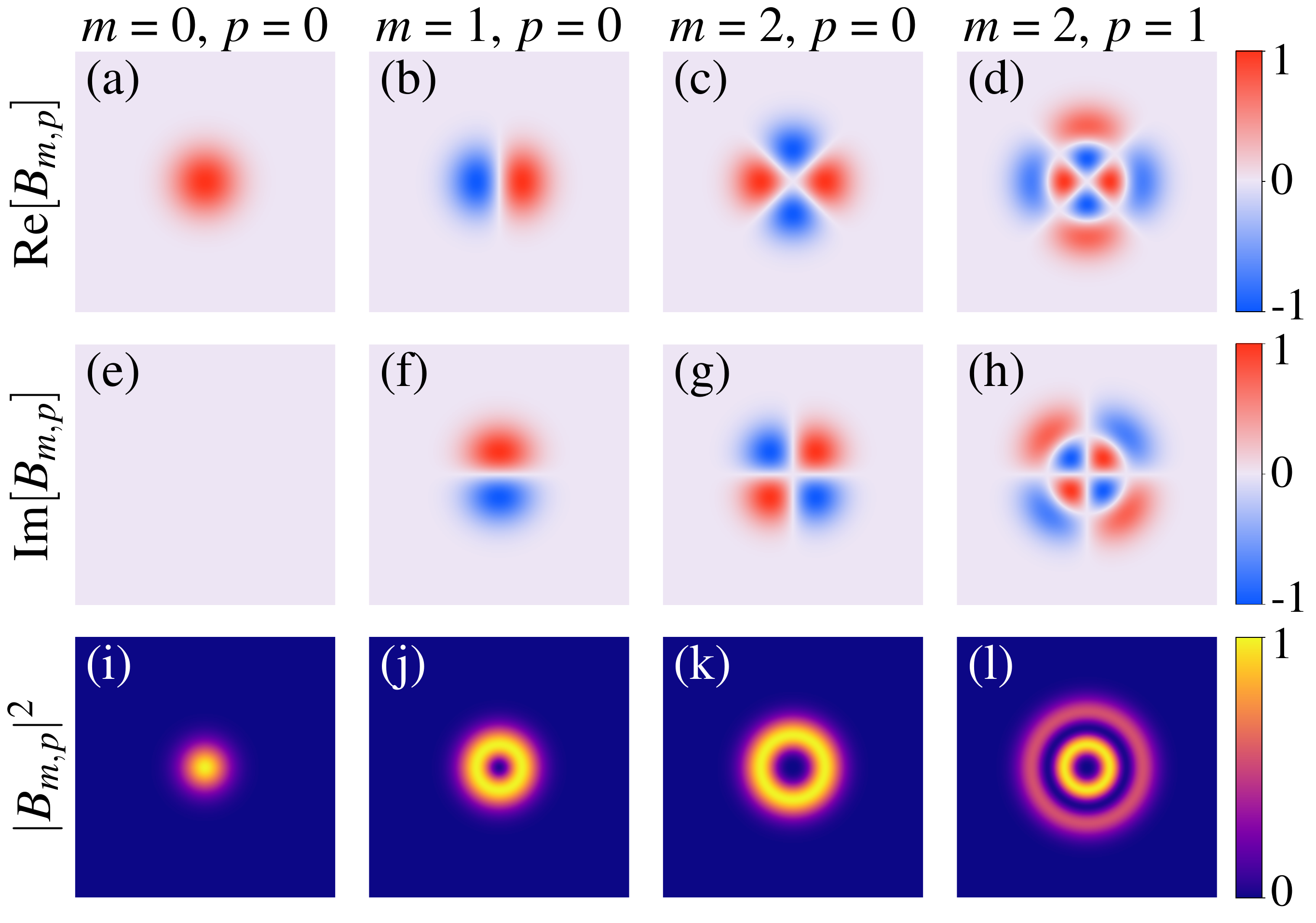}
	\caption{Normalized spatial profiles of LG modes at the focal plane ($z=0$). The first row, panels (a) to (d), shows the real part $\mathrm{Re}[B_{m,p}]$ of the OV magnetic field. The second row, panels (e) to (h), shows its imaginary part $\mathrm{Im}[B_{m,p}]$. The third row, panels (i) to (l), shows the corresponding intensity $|B_{m,p}|^2$. The columns correspond to $(m,p)=(0,0)$, $(1,0)$, $(2,0)$, and $(2,1)$, respectively. The fields and intensities are normalized to their respective peak values, with the color scales indicated by the colorbars.}
	\label{fig2}
\end{figure}

The spatial profiles of representative LG modes are shown in Fig.~\ref{fig2}. The first and second rows display the real and imaginary parts of the OV magnetic field, $\mathrm{Re}[B_{m,p}]$ and $\mathrm{Im}[B_{m,p}]$, respectively, while the third row shows the corresponding intensity $|B_{m,p}|^2$. For $(m,p)=(0,0)$, the field has a Gaussian profile with maximum intensity at the beam center [Fig.~\ref{fig2}(i)]. For $m\neq0$, the factor $(\rho/W)^{|m|}$ forces the field amplitude to vanish at $\rho=0$, producing an annular intensity structure [Figs.~\ref{fig2}(j)--(l)]. The azimuthal index $m$ sets the azimuthal phase winding of the mode: upon encircling the beam center, the phase accumulates $2\pi m$, and the real and imaginary field profiles exhibit $2|m|$ sign changes around the phase singularity. The radial index $p$ determines the number of radial nodes, resulting in $p+1$ concentric intensity maxima, as illustrated by the two-ring intensity profile of the $(m,p)=(2,1)$ mode in Fig.~\ref{fig2}(l). Modes with $p=0$ and $m=\pm 1$ are utilized to study the nucleation of single spin textures, whereas modes with $|m|>1$ or $p>0$ are used to generate multiple spin textures and clusters.

For our micromagnetic studies, we apply a pulsed OV to drive magnetization dynamics. The time-dependent magnetic field of the OV pulse is described by 
\begin{equation}
\mathbf{B}(\mathbf{r},t)
=
g(t)\,
\mathrm{Re}\!\left[
\mathbf{B}_{m,p}(\rho, \phi)
e^{-i\omega(t-\tau_1)}
\right].
\label{eq5_bt}
\end{equation}
Here $\omega=2\pi f$ is the angular frequency, $\mathbf{r}=(\rho, \phi)$ is the coordinate in the film plane, and the spatial profile of the vortex field $\mathbf{B}_{m,p}(\rho, \phi)$ is defined in Eq.~\eqref{eq:lg_mode}. The smooth time-dependent envelope $g(t)$ accounts for the consecutive turn-on and turn-off ramps:
\begin{equation}
g(t) = 
\begin{cases} 
\sin^2\left(\frac{\pi t}{2 \tau_{1}}\right), & 0 \le t < \tau_{1},\\
\cos^2\left(\frac{\pi (t - \tau_{1})}{2 \tau_{2}}\right), & \tau_{1} \le t < \tau_{1} + \tau_{2} ,\\
0, & \text{otherwise},
\end{cases}
\end{equation}
where $\tau_1$ and $\tau_2$ are the rise and fall times of the pulse, respectively, and the total pulse duration is
$t_{\mathrm{pulse}}=\tau_1+\tau_2$. The envelope reaches its maximum at $t=\tau_1$. 

The magnetization dynamics under the OV field is governed by the
Landau-Lifshitz-Gilbert equation,
\begin{equation}
\frac{d \mathbf{m}}{d t} = \gamma_0 \mathbf{H}_{\text{eff}} \times \mathbf{m} + \alpha \mathbf{m} \times \frac{d \mathbf{m}}{d t},
\end{equation}
where $\gamma_0$ is the gyromagnetic ratio, $\mathbf{H}_{\rm eff} = - (\mu_{0} M_{s})^{-1} \delta F[\mathbf{m}] / \delta \mathbf{m}$ is the effective magnetic field obtained from the free energy in Eq.~\eqref{eq1_fe}, with the time-dependent OV field $\mathbf{B}(\mathbf{r},t)$ defined in Eq.~\eqref{eq5_bt}, and $\alpha$
is the Gilbert damping constant. Unless stated otherwise, the simulations are performed for a film with lateral dimensions of $300\times300~\mathrm{nm}^2$ and a thickness of $2$~nm, discretized into $2\times2\times2~\mathrm{nm}^3$ cells with periodic boundary conditions imposed in the film plane. We use a beam waist $W=15$~nm, frequency $f=0.1$~THz, amplitude coefficient $B_0=10$~T, pulse fall time $\tau_2=0.5$~ps, and damping constant $\alpha=0.15$.

To study the generation of topological spin textures, we use the following three-stage writing protocol. Prior to the pulse, the magnetization in the film is relaxed into a uniform in-plane state. Then, during the pulse, the spatially inhomogeneous OV field drives rapid nonuniform precessional dynamics. The field amplitude reaches its maximum at $t=\tau_1$. After the field is switched off at $t_{\mathrm{pulse}}=\tau_1+\tau_2$, the subsequent magnetization evolution is governed by magnetic interactions and Gilbert damping, which relax the driven magnetization configuration either back to the uniform state or quench it into a nontrivial topological texture. Importantly, the optical parameters $(s,m,p)$ determine the symmetry and characteristic length scales of the transient magnetization pattern, while the subsequent magnetic relaxation stabilizes the resulting topological state as a single bimeron or antibimeron, or as a configuration of multiple spin textures. After full relaxation, the resulting magnetic state is characterized by the topological charge
\begin{equation}
Q=
\frac{1}{4\pi}
\int dxdy  \;
\mathbf{m}\cdot
\left(
\partial_x\mathbf{m}\times\partial_y\mathbf{m}
\right),
\label{eq:topological_charge}
\end{equation}
where for the in-plane systems studied below, nonzero values of $Q$ characterize single asymmetric bimerons or antibimerons, as well as clusters and other configurations composed of multiple spin textures. For configurations containing multiple textures, $Q$ represents the net topological charge and may have magnitude greater than one or vanish when textures with opposite charges compensate. Consequently, the value of $Q$ must be considered together with the spatial magnetization profile when characterizing multicomponent states. Note that the systems with the DMI types given by Eqs.~\eqref{eq:dmi_cnv} and \eqref{eq:dmi_d2d} can host energetically degenerate spin textures with opposite topological charges~\cite{Vorobyev2024}.

The distinction between asymmetric bimerons and asymmetric antibimerons is determined by the symmetry of the DMI that stabilizes these noncollinear textures. Both consist of a bound vortex-antivortex pair, but the DMI lifts the energetic equivalence of the two constituents by favoring one over the other. In a $C_{nv}$-symmetric system, the DMI favors the vortex constituent. The higher-energy antivortex consequently deforms into a crescent-shaped structure surrounding the vortex, thereby forming an asymmetric bimeron. Conversely, in a $D_{2d}$-symmetric system, the DMI favors the antivortex constituent~\cite{Vorobyev2024}. The vortex then deforms into a crescent-shaped structure surrounding the antivortex, producing an asymmetric antibimeron. Hereafter, asymmetric bimerons and antibimerons implicitly denote the textures stabilized by the DMI in Eq.~(\ref{eq:dmi_cnv}) and Eq.~(\ref{eq:dmi_d2d}), respectively.

\section{Generation of single in-plane spin textures}

Having established the micromagnetic framework and the writing protocol, we now investigate how the spatial profiles of LG beams, governed by their optical quantum numbers, nucleate in-plane topological spin textures in chiral FM films with $C_{nv}$ (or $D_{2d}$) symmetry. 
For the nucleation of single spin textures, we focus on beams with the radial index $p=0$ and OAM $m = 0, \pm 1$, considering both linear ($s=0$) and circular ($s=\pm1$) polarizations. 

\begin{figure}[h]  
	\centering
	\includegraphics[width=1\linewidth]{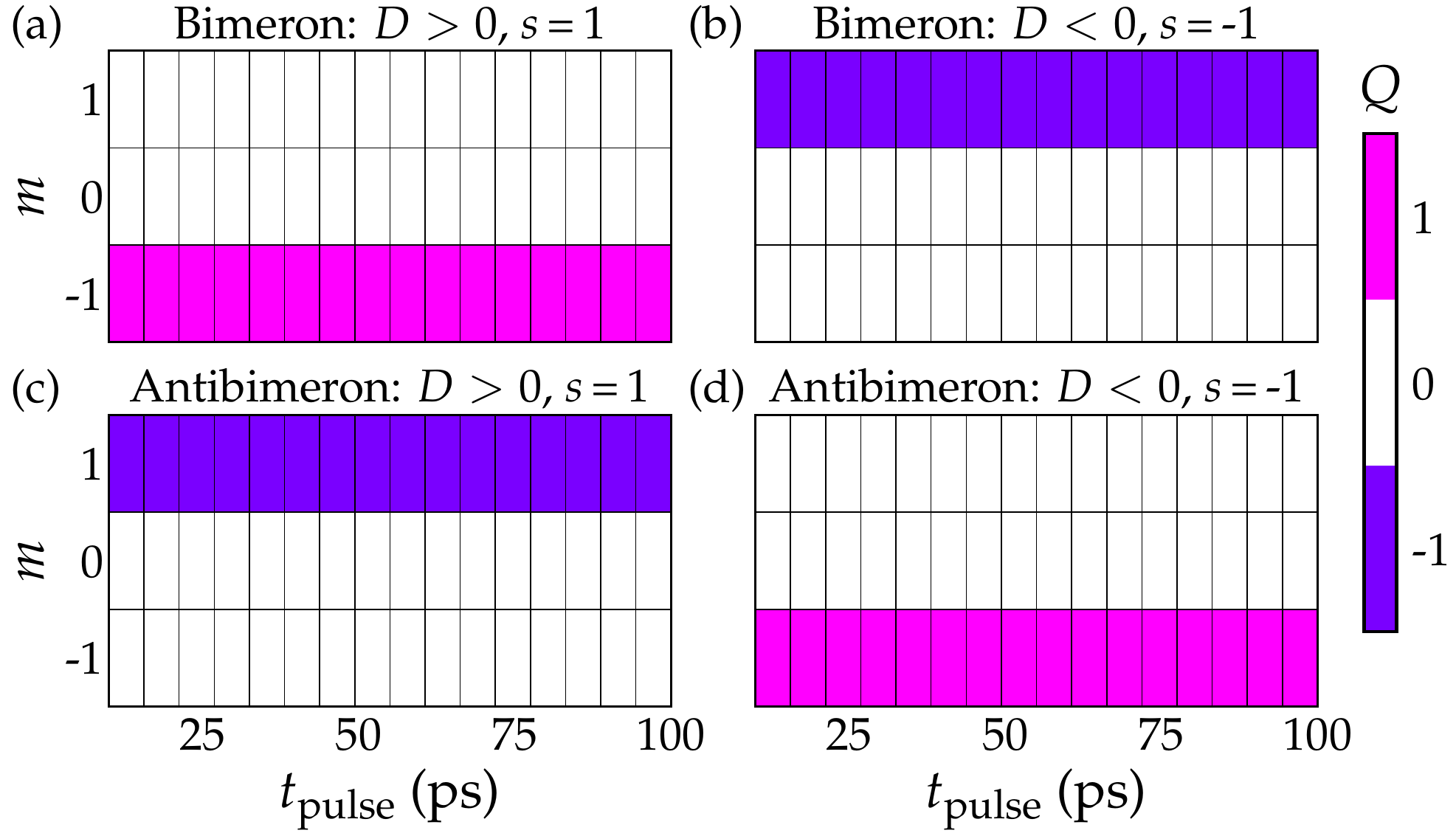}  
	\caption{Nucleation outcome diagrams for circularly polarized OVs. The color maps show the topological charge $Q$ as a function of the pulse duration $t_{\mathrm{pulse}}$ and the OAM $m$. Panels (a) and (b) correspond to the $C_{nv}$-symmetric system hosting asymmetric bimerons for (a) $D>0$ with beam polarization $s=1$, and (b) $D<0$ with $s=-1$. Panels (c) and (d) correspond to the $D_{2d}$-symmetric system hosting asymmetric antibimerons for (c) $D>0$ with $s=1$, and (d) $D<0$ with $s=-1$.}
	\label{fig3}
\end{figure}

Figure~\ref{fig3} summarizes the nucleation outcomes under circularly polarized OVs for systems with both DMI types and two signs of the DMI constant $D$. Specifically, the topological charge $Q$ of the nucleated magnetic texture is mapped as a function of the pulse duration $t_{\mathrm{pulse}}$ and the OAM $m$. A minimum pulse duration of $t_{\mathrm{pulse}} = 10$~ps (one optical cycle at $f = 0.1$~THz) is necessary to initiate nucleation, below which no topological transition occurs. For longer durations, the generation of the single texture remains robust across the entire investigated range up to 100~ps. For the $C_{nv}$-symmetric system, the stabilized configurations are asymmetric bimerons: an incident beam with $s = 1$ and $m = -1$ nucleates a state with $Q = 1$ when $D > 0$ [Fig.~\ref{fig3}(a)], whereas reversing the signs to $s = -1$ and $m = 1$ under $D < 0$ yields a state with $Q = -1$ [Fig.~\ref{fig3}(b)]. Conversely, the $D_{2d}$-symmetric system hosts asymmetric antibimerons: an incident beam with $m = 1$ and $s = 1$ stabilizes a state with $Q = -1$ when $D > 0$ [Fig.~\ref{fig3}(c)], whereas a beam with $s = -1$ and $m = -1$ yields a state with $Q = +1$ when $D < 0$ [Fig.~\ref{fig3}(d)]. Notably, in all cases, a standard Gaussian beam ($m = 0$) fails to induce a topological transition ($Q = 0$).

\begin{figure*}[t]  
	\centering
	\includegraphics[width=1\textwidth]{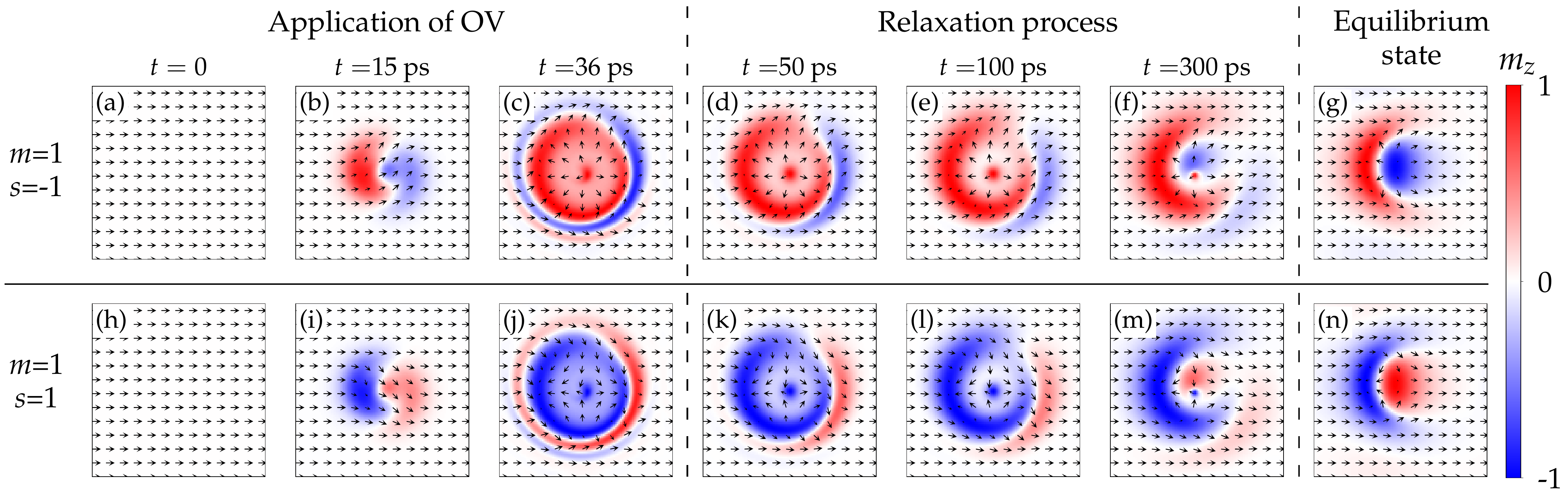}  
	\caption{Nucleation of single in-plane topological spin textures in a chiral FM driven by circularly polarized OVs. The OV pulse carries OAM index $m=1$ and has a total duration of $t_{\rm pulse}=36.5$~ps. The top row [panels (a)--(g)] corresponds to right-handed circular polarization ($s=-1$), leading to the nucleation of an asymmetric bimeron. The bottom row [panels (h)--(n)] corresponds to left-handed circular polarization ($s=1$), leading to the nucleation of an asymmetric antibimeron. Panels (a)--(c) and (h)--(j) show the magnetization dynamics during the OV pulse, while panels (d)--(f) and (k)--(m) show the subsequent relaxation after the pulse is switched off, with time reset to $t=0$ at the start of the relaxation stage. Panels (g) and (n) show the corresponding final fully relaxed states. The initial state is a uniform in-plane FM state with all spins aligned along the $+x$ direction. The color scale represents the out-of-plane component of the magnetization.}
	\label{fig4}
\end{figure*}

The magnetization dynamics illustrated in Fig.~\ref{fig4} elucidate the mechanism underlying the distinct nucleation outcomes for right- and left-handed circular polarizations summarized in Fig.~\ref{fig3}. For a fixed OAM $m=1$, the two circular polarization states ($s = \pm 1$) generate distinct nonuniform magnetization patterns during the OV pulse. As shown in Figs.~\ref{fig4}(a)--(c), an OV pulse with $s=-1$ induces a transient vortex pattern, which is required for the subsequent stabilization of an asymmetric bimeron [Figs.~\ref{fig4}(d)--(g)]. Conversely, in Figs.~\ref{fig4}(h)--(j), the pulse with $s=1$ excites an antivortex configuration that relaxes into an asymmetric antibimeron [Figs.~\ref{fig4}(k)--(n)]. These distinct pathways demonstrate that the transient vortex (antivortex) configuration directly determines the subsequent stabilization of bimeron (antibimeron), under the corresponding DMI type in the system. Interestingly, during the post-pulse relaxation stage, a reversal of the out-of-plane magnetization is observed within the central cores: the initially positive [$m_z > 0$, red in Fig.~\ref{fig4}(c)] vortex core flips to negative [$m_z < 0$, blue in Fig.~\ref{fig4}(g)] in the final bimeron state, while the initially negative antivortex core in Fig.~\ref{fig4}(j) switches to positive in Fig.~\ref{fig4}(n). Finally, as shown in Fig.~\ref{fig4}, choosing a beam waist $W$ comparable to the characteristic size of the target spin texture is important. This spatial matching enables the OV pulse to generate a localized transient configuration on the appropriate length scale, allowing it to relax efficiently into a stable asymmetric spin texture.

Furthermore, the nucleation outcomes shown in Fig.~\ref{fig3} reveal an OAM-dependent response governed by the compatibility between the chirality of the OV and the intrinsic chirality of the FM film. The OAM-dependent OV field induces a twist in the magnetization, with the sign of the OAM ($m=\pm 1$) determining the direction of this twist. The sign of the DMI constant $D$ sets the intrinsic chirality of the film by energetically favoring a particular rotational sense of the magnetization. Thus, in the presence of DMI, the OV-induced twist either matches or competes with the film’s intrinsic chirality. When the two chiralities are compatible, the transient magnetization pattern can relax into a topological spin texture. In contrast, opposing chiralities prevent stabilization and suppress the topological transition. Therefore, the OAM-sign dependence of the nucleation outcomes can be understood as arising from the competition between optical and material chiralities.

\begin{figure}[h]  
	\centering
	\includegraphics[width=1\linewidth]{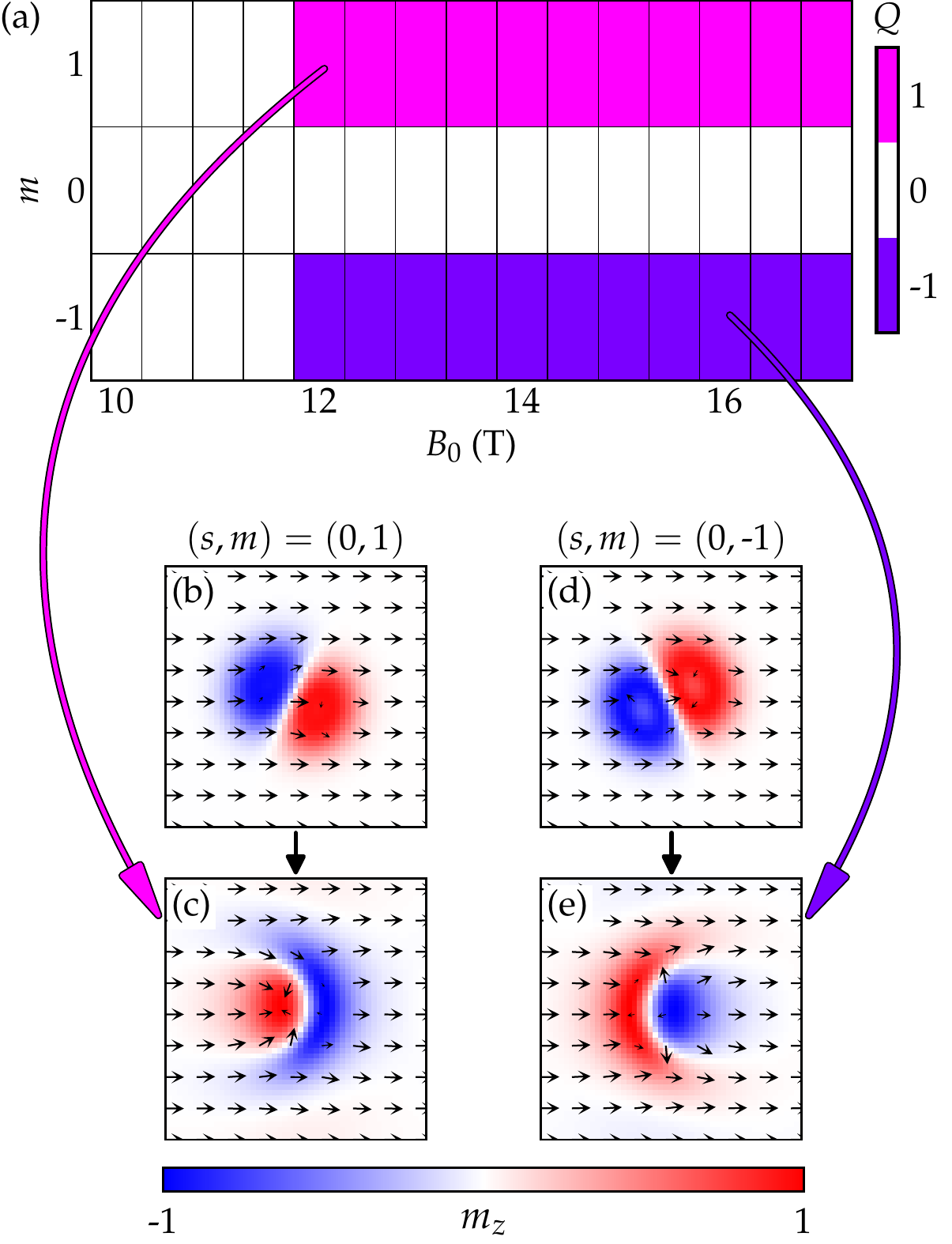}  
	\caption{(a) Nucleation outcome diagram for linearly polarized OVs. The color map shows the topological charge $Q$ as a function of the OV magnetic-field amplitude coefficient $B_0$ and OAM $m$. (b)--(e) Nucleation of asymmetric bimerons. Panels (b) and (d) show the magnetization states at $t=5$~ps during the OV pulse for $m=1$ and $m=-1$ at $B_0=12$~T and $B_0=16$~T, respectively. The corresponding equilibrium states after relaxation are shown in (c), with $Q=1$, and in (e), with $Q=-1$. The color in panels (b)--(e) represents the out-of-plane magnetization component.}
	\label{fig5}
\end{figure}

We next consider the nucleation of topological spin textures induced by linearly polarized OVs with $s = 0$, where the polarization is chosen along the $y$-direction. Unlike the circularly polarized case, in which  $B_0=10$~T was sufficient to induce nucleation, the same value of $B_0$ under linear polarization does not produce stable topological textures after relaxation. Therefore, instead of varying the pulse duration, we sweep $B_0$ to determine the threshold for stable nucleation. Figure~\ref{fig5}(a) summarizes the nucleation outcomes for $m=0$ and $m=\pm1$ at a fixed pulse duration of $t_{\mathrm{pulse}}=10$~ps. The results show that successful nucleation requires $B_0\simeq12$~T. This higher threshold arises because a linearly polarized field can be viewed as an equal superposition of the two circularly polarized states with $s=\pm1$. Consequently, the net SAM vanishes, suppressing the transient magnetic vortex (or antivortex) that mediates nucleation in the corresponding circularly polarized cases. Instead, the transient states formed during the pulse exhibit magnetization patterns lacking such well-defined structures [see Figs.~\ref{fig5}(b) and \ref{fig5}(d)]. Below this threshold, the induced transient magnetization configuration is insufficient to relax into a stable topological texture. Above $B_0\simeq12$~T, the relaxed state produced by a standard Gaussian beam ($m=0$) remains topologically trivial with $Q=0$, whereas vortex beams with $m=\pm1$ successfully nucleate corresponding topological spin textures.

Crucially, under linear polarization, the topological charge of the nucleated texture directly follows the OAM through the relation $Q=m$. An incident beam with $m=+1$ produces an equilibrium texture with $Q=+1$ [Fig.~\ref{fig5}(c)], whereas a beam with $m=-1$ produces a texture with $Q=-1$ [Fig.~\ref{fig5}(e)]. This correspondence is independent of the DMI type. The DMI only determines whether the resulting equilibrium state is an asymmetric bimeron or antibimeron, but it does not change its topological charge. Thus, linear polarization enables direct one-to-one encoding of the beam’s OAM into the topological charge of the nucleated spin texture.

\section{Generation of in-plane clusters and multiple spin textures}

Asymmetric spin textures, including bimerons and antibimerons, can be stabilized not only as isolated objects but also as clusters~\cite{Li20, Vorobyev2026}. This clustering tendency arises because, unlike symmetric bimerons~\cite{Gobel2019}, asymmetric in-plane textures with the same topological charge exhibit an attractive interaction along the direction perpendicular to the easy axis. This attraction promotes the formation of clusters with a broad range of total topological charges $Q$. In the following, we show that OV beams with higher OAM ($|m|>1$) or nonzero radial indices ($p>0$) can nucleate such clusters, as well as other multicomponent states in which several spin textures coexist.

\begin{figure}[h]
	\centering
	\includegraphics[width=1\linewidth]{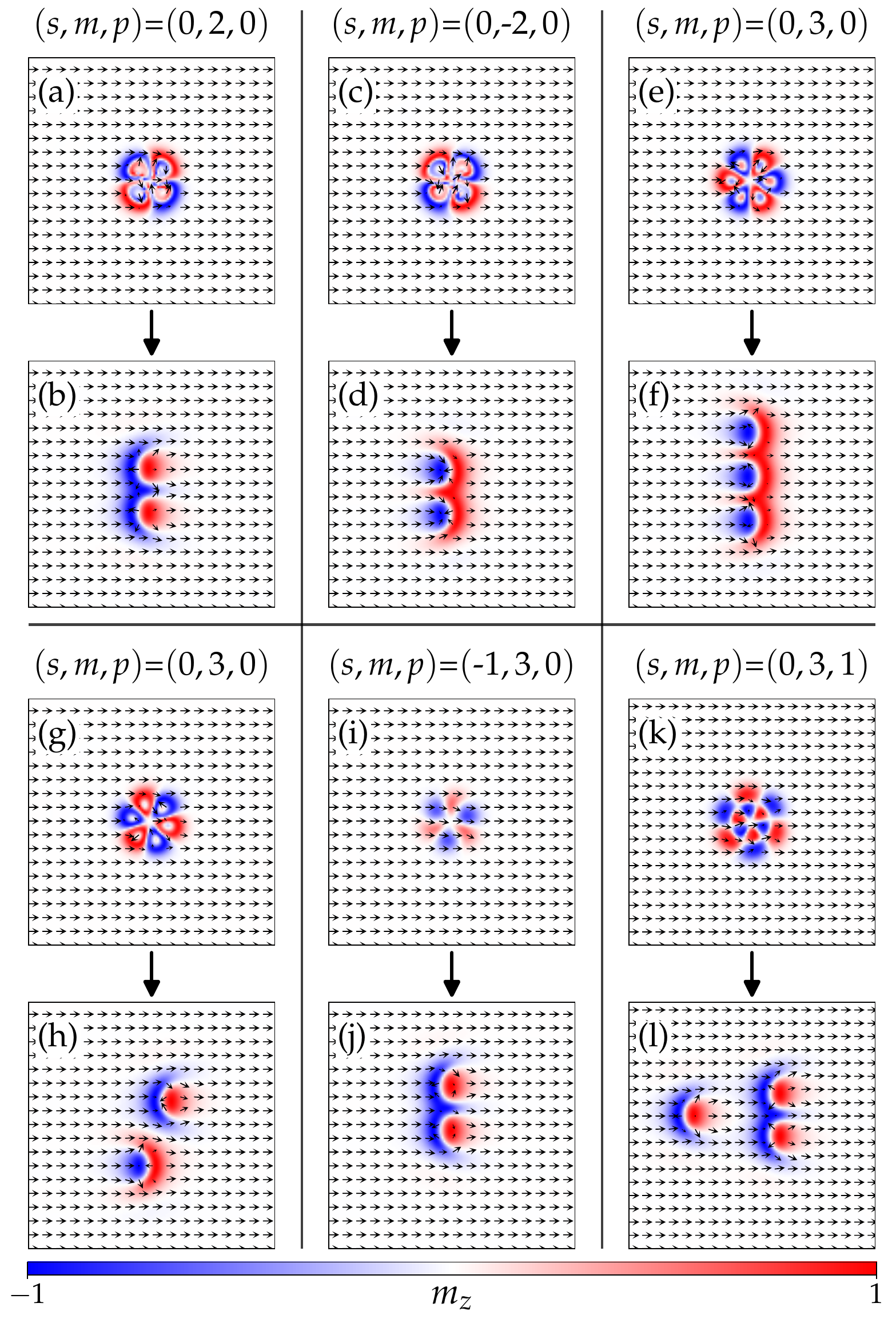}
	\caption{Nucleation of multiple in-plane topological spin textures induced by OVs. For each LG mode $(s,m,p)$, the upper and lower panels show the magnetization state at $t=10$~ps and the corresponding equilibrium state after relaxation, respectively. Results are shown for the following modes: (a,b) $(0,2,0)$, which nucleates a bimeron cluster with $Q=2$; (c,d) $(0,-2,0)$, which nucleates a bimeron cluster with $Q=-2$; (e,f) $(0,3,0)$, which nucleates an antibimeron cluster with $Q=3$; (g,h) $(0,3,0)$, which nucleates isolated antibimerons with $Q=+1$ and $Q=-1$; (i,j) $(-1,3,0)$, which nucleates an antibimeron cluster with $Q=-2$; and (k,l) $(0,3,1)$, which nucleates a bimeron cluster with $Q=2$ and an isolated bimeron with $Q=1$. The color represents the out-of-plane component of the magnetization.}
	\label{fig6}
\end{figure}

In Fig.~\ref{fig6}, we demonstrate the nucleation of various topological states using OV field profiles with OAM $|m|>1$. Each case is labeled by the optical quantum numbers $(s,m,p)$. The upper panels show the magnetization patterns generated during the pulse, whereas the lower panels show the corresponding equilibrium states after relaxation. The black arrows indicate the evolution from the transient magnetization state during the pulse to the final relaxed state. We first consider cases in which the OAM of the OV is encoded into the magnetic film as the total topological charge of the nucleated cluster, i.e., $Q=m$. For example, an OV profile with $(s,m,p)=(0,2,0)$, $B_0=18$~T, and $t_{\mathrm{pulse}}=10$~ps nucleates a bimeron cluster with $Q=2$ [Figs.~\ref{fig6}(a,b)]. Conversely, the $(0,-2,0)$ OV profile with $B_0=18.5$~T and $t_{\mathrm{pulse}}=10$~ps nucleates a bimeron cluster with $Q=-2$ [Figs.~\ref{fig6}(c,d)]. This topological encoding also extends to larger OAM values: for the $(0,3,0)$ OV profile with $B_0=22$~T and $t_{\mathrm{pulse}}=18.5$~ps nucleates an antibimeron cluster with $Q=3$ [Figs.~\ref{fig6}(e,f)]. These examples show that, by tuning $B_0$ and pulse duration $t_{\mathrm{pulse}}$, OVs with $|m|>1$ can be tailored to nucleate clusters whose total topological charge is exactly equal to the beam OAM.

The need for such precise parameter tuning originates from the spatial structure of OV fields. For beams with $|m|>1$, the annular field profile splits azimuthally into $2|m|$ petals of alternating sign and equal magnitude [see, e.g., Fig.~\ref{fig2}(c)]. Each petal acts as a localized field maximum, creating multiple competing nucleation sites. At a fixed beam waist $W$, increasing the OAM index produces sharper spatial gradients within the same area and reduces the size of individual petals below the characteristic length scale of the target spin textures. This length-scale mismatch can suppress nucleation altogether or prevent cluster formation, instead producing decoupled spin textures. An example of this breakdown is shown in Figs.~\ref{fig6}(g,h) for the same $(0,3,0)$ mode used in Figs.~\ref{fig6}(e,f), but with a different pulse configuration: $B_0=30$~T and $t_{\mathrm{pulse}}=24.5$~ps. In this case, the OV field nucleates a pair of isolated antibimerons with opposite topological charges, $Q=+1$ and $Q=-1$. The appearance of textures with opposite topological charges reflects complex transient magnetization dynamics, including possible annihilation events, which complicate deterministic control over the final equilibrium state.

Moreover, the clusters tend to align along the $y$-axis [see, e.g., Fig.~\ref{fig6}(f)], whereas OV fields combine a cylindrically symmetric radial envelope with an OAM-controlled azimuthal phase winding [see, e.g., Fig.~\ref{fig2}(c)]. This symmetry mismatch between the preferred cluster orientation and the structure of the driving field makes the final magnetization state sensitive to additional optical degrees of freedom. For example, changing the SAM and using the $(-1,3,0)$ mode with $B_0=30$~T and $t_{\mathrm{pulse}}=54.5$~ps nucleates an antibimeron cluster with $Q=-2$ [Figs.~\ref{fig6}(i,j)]. Alternatively, introducing a radial node by choosing the $(0,3,1)$ mode with $B_0=22$~T and $t_{\mathrm{pulse}}=30.5$~ps produces a concentric multi-ring field profile. Upon relaxation, this complex spatial pattern evolves into a coexistence state consisting of a $Q=2$ bimeron cluster and a single $Q=1$ bimeron [Figs.~\ref{fig6}(k,l)]. Thus, although one-to-one encoding of OAM into topological charge is no longer straightforwardly preserved in this regime, the expanded optical parameter space provides access to a rich variety of topological states.

\section{Conclusion}

In this work, we have demonstrated a rich and promising approach at the intersection of structured photonics~\cite{Forbes2021} and spintronics for creating in-plane topological spin textures in chiral magnets. Using micromagnetic simulations, we have shown that the spatial structure of OVs provides a direct route for encoding topological charge into these systems. By leveraging control over the optical quantum numbers of the incident LG beams, namely the SAM, OAM, and radial index, we have generated a broad variety of nontrivial topological spin textures, spanning from isolated asymmetric bimerons and antibimerons to their clusters.

For OVs with $|m|=1$, we have achieved the deterministic writing of single asymmetric bimerons or antibimerons, stabilized by the respective DMI allowed under $C_{nv}$ or $D_{2d}$ symmetries. For circularly polarized OVs, the encoding of the beam's OAM into the topological charge $Q$ requires matching the SAM and OAM with the magnetic chirality selected by the DMI type and sign. For linearly polarized OVs, the absence of net SAM simplifies the nucleation process, enabling direct encoding in which the topological charge $Q$ is set by the sign and magnitude of the OAM, independent of the DMI type. For OVs with $|m|>1$, clusters of spin textures can be generated under appropriate conditions. Fine tuning of the OV parameters allows a direct encoding, where the topological charge exactly matches the OAM, $Q=m$. Introducing a nonzero radial index $p>0$ or nonzero SAM further enriches this encoding, producing spin textures with diverse topological charges. For combinations of $|m|>1$ and $p>0$, the OV spatial profile may be incompatible with the preferred geometry of the clusters, which tend to elongate along the $y$-direction. This mismatch can give rise to distinct topological states, including configurations in which clusters coexist with isolated textures.

In our micromagnetic framework, the optical stimulus is implemented as a pulsed sub-THz optical vortex with a nanoscale beam waist. This choice provides a proof-of-principle implementation in which the beam profile is matched to the characteristic size of the target spin textures. Experimentally, such tight spatial confinement below the far-field diffraction limit may be achieved using near-field optical techniques, including near-field scanning optical microscopy~\cite{Prinz2023, Guo2024}, plasmonic antennas~\cite{Reinier2014}, or near-field photoconductive antennas~\cite{Tuniz2023}. Furthermore, specialized metasurfaces provide another route for generating topologically structured THz fields~\cite{Niu2026}. Overall, our findings establish structured light as a versatile tool for encoding optical quantum numbers into in-plane spin textures in chiral magnets, offering a pathway toward topology-selective writing for spintronic applications. 

\section*{Acknowledgments} 
O.A.T. acknowledges support from the Australian Research Council (Grant No. DP240101062), NCMAS grant, International Excellence Fellowship of KIT, and visiting program of ICC-IMR, Tohoku University (Japan).

\bibliography{optical_vortex}

\end{document}